\documentclass[reprint,aps,prl,twocolumn,superscriptaddress,showpacs,preprintnumbers,amsmath,amssymb,floatfix]{revtex4-1}

\usepackage{graphicx}
\usepackage[mathlines]{lineno}
\usepackage{xspace} 
\usepackage{wasysym}

\newcommand{\gev}{GeV/c$^2$\xspace}
\newcommand{\tev}{TeV/c$^2$\xspace}
\newcommand{\ba}{$^{133}$Ba\xspace}
\newcommand{\cf}{$^{252}$Cf\xspace}

\begin{document}
\title{Results from a Low-Energy Analysis of the CDMS~II Germanium Data}


\affiliation{Division of Physics, Mathematics \& Astronomy, California Institute of Technology, Pasadena, CA 91125, USA} 
\affiliation{Department of Physics, Case Western Reserve University, Cleveland, OH  44106, USA}
\affiliation{Fermi National Accelerator Laboratory, Batavia, IL 60510, USA}
\affiliation{Lawrence Berkeley National Laboratory, Berkeley, CA 94720, USA}
\affiliation{Department of Physics, Massachusetts Institute of Technology, Cambridge, MA 02139, USA}
\affiliation{Department of Physics, Queen's University, Kingston, ON, Canada, K7L 3N6}
\affiliation{SLAC National Accelerator Laboratory/KIPAC, Menlo Park, CA 94025, USA}
\affiliation{Department of Physics, St.\,Olaf College, Northfield, MN 55057 USA}
\affiliation{Department of Physics, Santa Clara University, Santa Clara, CA 95053, USA}
\affiliation{Department of Physics, Southern Methodist University, Dallas, TX 75275, USA}
\affiliation{Department of Physics, Stanford University, Stanford, CA 94305, USA}
\affiliation{Department of Physics, Syracuse University, Syracuse, NY 13244, USA}
\affiliation{Department of Physics, Texas A \& M University, College Station, TX 77843, USA}
\affiliation{Department of Physics, University of California, Berkeley, CA 94720, USA}
\affiliation{Department of Physics, University of California, Santa Barbara, CA 93106, USA}
\affiliation{Departments of Phys. \& Elec. Engr., University of Colorado Denver, Denver, CO 80217, USA}
\affiliation{Department of Physics, University of Florida, Gainesville, FL 32611, USA}
\affiliation{School of Physics \& Astronomy, University of Minnesota, Minneapolis, MN 55455, USA}
\affiliation{Physics Institute, University of Z\"{u}rich, Winterthurerstr. 190, CH-8057, Switzerland}

\author{Z.~Ahmed} \affiliation{Division of Physics, Mathematics \& Astronomy, California Institute of Technology, Pasadena, CA 91125, USA} 
\author{D.S.~Akerib} \affiliation{Department of Physics, Case Western Reserve University, Cleveland, OH  44106, USA} 
\author{S.~Arrenberg} \affiliation{Physics Institute, University of Z\"{u}rich, Winterthurerstr. 190, CH-8057, Switzerland}
\author{C.N.~Bailey} \affiliation{Department of Physics, Case Western Reserve University, Cleveland, OH  44106, USA} 
\author{D.~Balakishiyeva} \affiliation{Department of Physics, University of Florida, Gainesville, FL 32611, USA} 
\author{L.~Baudis} \affiliation{Physics Institute, University of Z\"{u}rich, Winterthurerstr. 190, CH-8057, Switzerland}
\author{D.A.~Bauer} \affiliation{Fermi National Accelerator Laboratory, Batavia, IL 60510, USA} 
\author{P.L.~Brink} \affiliation{SLAC National Accelerator Laboratory/KIPAC, Menlo Park, CA 94025, USA}
\author{T.~Bruch} \affiliation{Physics Institute, University of Z\"{u}rich, Winterthurerstr. 190, CH-8057, Switzerland}
\author{R.~Bunker} \affiliation{Department of Physics, University of California, Santa Barbara, CA 93106, USA} 
\author{B.~Cabrera} \affiliation{Department of Physics, Stanford University, Stanford, CA 94305, USA} 
\author{D.O.~Caldwell} \affiliation{Department of Physics, University of California, Santa Barbara, CA 93106, USA} 
\author{J.~Cooley} \affiliation{Department of Physics, Southern Methodist University, Dallas, TX 75275, USA} 
\author{E.~do~Couto~e~Silva} \affiliation{SLAC National Accelerator Laboratory/KIPAC, Menlo Park, CA 94025, USA} 
\author{P.~Cushman} \affiliation{School of Physics \& Astronomy, University of Minnesota, Minneapolis, MN 55455, USA} 
\author{M.~Daal} \affiliation{Department of Physics, University of California, Berkeley, CA 94720, USA} 
\author{F.~DeJongh} \affiliation{Fermi National Accelerator Laboratory, Batavia, IL 60510, USA} 
\author{P.~Di~Stefano} \affiliation{Department of Physics, Queen's University, Kingston, ON, Canada, K7L 3N6}
\author{M.R.~Dragowsky} \affiliation{Department of Physics, Case Western Reserve University, Cleveland, OH  44106, USA} 
\author{L.~Duong} \affiliation{School of Physics \& Astronomy, University of Minnesota, Minneapolis, MN 55455, USA} 
\author{S. Fallows}\affiliation{School of Physics \& Astronomy, University of Minnesota, Minneapolis, MN 55455, USA} 
\author{E.~Figueroa-Feliciano} \affiliation{Department of Physics, Massachusetts Institute of Technology, Cambridge, MA 02139, USA} 
\author{J.~Filippini} \affiliation{Division of Physics, Mathematics \& Astronomy, California Institute of Technology, Pasadena, CA 91125, USA} 
\author{J.~Fox} \affiliation{Department of Physics, Queen's University, Kingston, ON, Canada, K7L 3N6}
\author{M.~Fritts} \affiliation{School of Physics \& Astronomy, University of Minnesota, Minneapolis, MN 55455, USA} 
\author{S.R.~Golwala} \affiliation{Division of Physics, Mathematics \& Astronomy, California Institute of Technology, Pasadena, CA 91125, USA} 
\author{J.~Hall} \affiliation{Fermi National Accelerator Laboratory, Batavia, IL 60510, USA} 
\author{R.~Hennings-Yeomans} \affiliation{Department of Physics, Case Western Reserve University, Cleveland, OH  44106, USA} 
\author{S.A.~Hertel} \affiliation{Department of Physics, Massachusetts Institute of Technology, Cambridge, MA 02139, USA} 
\author{D.~Holmgren} \affiliation{Fermi National Accelerator Laboratory, Batavia, IL 60510, USA} 
\author{L.~Hsu} \affiliation{Fermi National Accelerator Laboratory, Batavia, IL 60510, USA} 
\author{M.E.~Huber} \affiliation{Departments of Phys. \& Elec. Engr., University of Colorado Denver, Denver, CO 80217, USA}
\author{O.~Kamaev}\affiliation{School of Physics \& Astronomy, University of Minnesota, Minneapolis, MN 55455, USA} 
\author{M.~Kiveni} \affiliation{Department of Physics, Syracuse University, Syracuse, NY 13244, USA} 
\author{M.~Kos} \affiliation{Department of Physics, Syracuse University, Syracuse, NY 13244, USA} 
\author{S.W.~Leman} \affiliation{Department of Physics, Massachusetts Institute of Technology, Cambridge, MA 02139, USA} 
\author{S.~Liu} \affiliation{Department of Physics, Queen's University, Kingston, ON, Canada, K7L 3N6}
\author{R.~Mahapatra} \affiliation{Department of Physics, Texas A \& M University, College Station, TX 77843, USA} 
\author{V.~Mandic} \affiliation{School of Physics \& Astronomy, University of Minnesota, Minneapolis, MN 55455, USA} 
\author{K.A.~McCarthy} \affiliation{Department of Physics, Massachusetts Institute of Technology, Cambridge, MA 02139, USA} 
\author{N.~Mirabolfathi} \affiliation{Department of Physics, University of California, Berkeley, CA 94720, USA} 
\author{D.~Moore}  \email{Corresponding author: davidm@caltech.edu} \affiliation{Division of Physics, Mathematics \& Astronomy, California Institute of Technology, Pasadena, CA 91125, USA}
\author{H.~Nelson} \affiliation{Department of Physics, University of California, Santa Barbara, CA 93106, USA} 
\author{R.W.~Ogburn}\affiliation{Department of Physics, Stanford University, Stanford, CA 94305, USA} 
\author{A.~Phipps}\affiliation{Department of Physics, University of California, Berkeley, CA 94720, USA} 
\author{M.~Pyle} \affiliation{Department of Physics, Stanford University, Stanford, CA 94305, USA} 
\author{X.~Qiu} \affiliation{School of Physics \& Astronomy, University of Minnesota, Minneapolis, MN 55455, USA} 
\author{E.~Ramberg} \affiliation{Fermi National Accelerator Laboratory, Batavia, IL 60510, USA} 
\author{W.~Rau} \affiliation{Department of Physics, Queen's University, Kingston, ON, Canada, K7L 3N6}
\author{A.~Reisetter} \affiliation{School of Physics \& Astronomy, University of Minnesota, Minneapolis, MN 55455, USA} \affiliation{Department of Physics, St.\,Olaf College, Northfield, MN 55057 USA}
\author{R.~Resch} \affiliation{SLAC National Accelerator Laboratory/KIPAC, Menlo Park, CA 94025, USA}  
\author{T.~Saab} \affiliation{Department of Physics, University of Florida, Gainesville, FL 32611, USA}
\author{B.~Sadoulet} \affiliation{Lawrence Berkeley National Laboratory, Berkeley, CA 94720, USA} \affiliation{Department of Physics, University of California, Berkeley, CA 94720, USA}
\author{J.~Sander} \affiliation{Department of Physics, University of California, Santa Barbara, CA 93106, USA} 
\author{R.W.~Schnee} \affiliation{Department of Physics, Syracuse University, Syracuse, NY 13244, USA} 
\author{D.N.~Seitz} \affiliation{Department of Physics, University of California, Berkeley, CA 94720, USA} 
\author{B.~Serfass} \affiliation{Department of Physics, University of California, Berkeley, CA 94720, USA} 
\author{K.M.~Sundqvist} \affiliation{Department of Physics, University of California, Berkeley, CA 94720, USA} 
\author{M.~Tarka}\affiliation{Physics Institute, University of Z\"{u}rich, Winterthurerstr. 190, CH-8057, Switzerland}
\author{P.~Wikus} \affiliation{Department of Physics, Massachusetts Institute of Technology, Cambridge, MA 02139, USA} 
\author{S.~Yellin} \affiliation{Department of Physics, Stanford University, Stanford, CA 94305, USA} \affiliation{Department of Physics, University of California, Santa Barbara, CA 93106, USA}
\author{J.~Yoo} \affiliation{Fermi National Accelerator Laboratory, Batavia, IL 60510, USA} 
\author{B.A.~Young} \affiliation{Department of Physics, Santa Clara University, Santa Clara, CA 95053, USA} 
\author{J.~Zhang}\affiliation{School of Physics \& Astronomy, University of Minnesota, Minneapolis, MN 55455, USA} 

\collaboration{CDMS Collaboration}

\noaffiliation

\begin{abstract}
We report results from a reanalysis of data from the Cryogenic Dark Matter Search (CDMS~II) experiment at the Soudan Underground Laboratory. Data taken between October 2006 and September 2008 using eight germanium detectors are reanalyzed with a lowered, 2~keV recoil-energy threshold, to give increased sensitivity to interactions from Weakly Interacting Massive Particles (WIMPs) with masses below $\sim$10~GeV/c$^2$. This analysis provides stronger constraints than previous CDMS~II results for WIMP masses below 9~GeV/c$^2$ and excludes parameter space associated with possible low-mass WIMP signals from the DAMA/LIBRA and CoGeNT experiments. 
\end{abstract}


\maketitle

 
A convergence of astrophysical observations indicates that $\sim$80\% of the matter in the universe is in the form of non-baryonic, non-luminous dark matter~\cite{Bertone:2004pz}. Weakly Interacting Massive Particles (WIMPs)~\cite{Steigman:1984ac}, with masses from a few \gev to a few \tev, form a well-motivated class of candidates for this dark matter~\cite{Bertone:2004pz,Jungman:1995df}. If WIMPs account for the dark matter, they may be detectable through their elastic scattering with nuclei in terrestrial detectors~\cite{Goodman:1984dc,*Gaitskell:2004gd}.

Although many models of physics beyond the Standard Model provide WIMP candidates, supersymmetric (SUSY) models where the lightest superpartner is a cosmologically stable WIMP are among the most popular~\cite{Bertone:2004pz,Jungman:1995df}. In the Minimal Supersymmetric Standard Model (MSSM), WIMPs with masses $\lesssim$40~\gev are generally disfavored by accelerator constraints (e.g.,~\cite{Baltz:2004aw,*Heister:2004}). Interest in lower-mass WIMPs has been renewed by recent results from the DAMA/LIBRA~\cite{Bernabei:2008yi,*Bernabei:2010} and CoGeNT~\cite{Aalseth:2010vx} experiments, which have been interpreted in terms of elastic scatters from a WIMP with mass $\sim$10~\gev and cross-section $\sim$10$^{-40}$~cm$^2$~\cite{Hooper:2010ly,Fitzpatrick:2010ve,*Chang:2010,*Andreas:2010qf}. Although it is difficult to accommodate a WIMP with these properties in the MSSM~\cite{Kuflik:2010zr,*Feldman:2010,*Hooper:2003}, alternate models avoid existing constraints (e.g.,~\cite{Bottino:2003cz,*Draper:2010,*Albornoz:2010,*Belikov:2010fk,*Belanger:2004,*Feng:2008,*Andreas:2008,*Zurek:2009}).

The CDMS~II experiment attempts to identify nuclear recoils from WIMPs in an array of particle detectors by measuring both the ionization and non-equilibrium phonons created by each particle interaction. Backgrounds can be rejected on an event-by-event basis since they primarily scatter from electrons in the detector, depositing significantly more ionization than a nuclear recoil of the same energy. Previous analyses of CDMS~II data~\cite{Ahmed:2008eu,*CDMSScience:2010} imposed a recoil-energy threshold of 10~keV to maintain sufficient rejection of electron recoils that only $\sim$0.5 background events would be expected in the signal region.  At lower energies, the discrimination between nuclear and electron recoils degrades, leading to higher expected backgrounds.  Since WIMPs with masses $<$10~\gev primarily produce $<$10~keV recoils, this analysis lowers the recoil-energy threshold to 2~keV, comparable to the hardware trigger threshold.  This lower energy threshold increases sensitivity to low-mass WIMPs at the cost of significant acceptance of backgrounds.

The data analyzed here were collected using all 30 Z-sensitive Ionization and Phonon (ZIP) detectors installed at the Soudan Underground Laboratory~\cite{Akerib:2005zy,Ahmed:2008eu,*CDMSScience:2010}. The detector array consisted of 19 Ge ($\sim$230~g each) and 11 Si ($\sim$105~g each) detectors, each a disk $\sim$10~mm thick and 76~mm in diameter. Each detector was instrumented with four phonon sensors on one face and two concentric charge electrodes on the opposite face.  A small electric field (3\textendash4 V/cm) was applied across the detectors to extract charge carriers created by particle interactions. The detectors were arranged in five ``towers,'' and are identified by their tower number (T1\textendash T5) and by their ordering within the tower (Z1\textendash Z6). The entire array was cooled to $\lesssim$50 mK and surrounded by passive lead and polyethylene shielding. An outer plastic scintillator veto was used to identify showers containing cosmogenic muons which were not shielded by the rock overburden above the Soudan laboratory (2090 meters water equivalent).  

The data were taken during six data runs from October 2006 to September 2008~\cite{Ahmed:2008eu,*CDMSScience:2010}.  Only the eight Ge detectors with the lowest trigger thresholds were used to identify WIMP candidate events since they have the best expected sensitivity to WIMPs with masses from 5\textendash10~\gev. All 30 detectors were used to veto events that deposited energy in multiple detectors. 

Each detector was monitored throughout the data runs and periods of abnormal detector performance were removed~\cite{Ahmed:2008eu,*CDMSScience:2010}. Data taken within 20 days following exposure of the detectors to a neutron calibration source were removed to reduce low-energy electron-recoil backgrounds due to activation of the detectors.  The data were randomly divided into two subsets before defining selection criteria at low energy. One subset, consisting of one quarter of the data (the ``open'' data), was reserved to study backgrounds at low energy and was not used to calculate exclusion limits. The remaining subset totaled 241 kg days raw exposure, after removing the bad data periods described above.

The detector response to electron and nuclear recoils was calibrated by regular exposures of the detectors to $\gamma$-ray (\ba) and neutron (\cf) sources. The ionization energy scale was initially calibrated using the 356~keV line from the \ba source. The phonon energy was then calibrated by normalizing the phonon-based recoil energy for electron recoils to their mean ionization energy. In contrast to previous analyses, a position-dependent calibration was not applied since position-dependent variations in the reconstructed phonon energies are less significant than noise fluctuations at low energies. Using the observed positions of the 1.3~keV and 10.4~keV activation lines produced from exposure of the Ge detectors to the \cf source, a small rescaling ($\sim$4\%) was applied to ensure that the recoil energy scale for electron recoils was not underestimated at the 90\% confidence level.  For nuclear recoils, the recoil energy was reconstructed from the measured phonon energy alone by subtracting the Neganov-Luke phonon contribution~\cite{Neganov:1985,*Luke:1988} corresponding to the mean ionization measured for nuclear recoils of the same phonon energy from the \cf source.  The ratio of ionization to recoil energy (``ionization yield'') for nuclear recoils was measured down to $\sim$4~keV, below which a power-law extrapolation was used.  

Candidate events were required to pass basic reconstruction quality cuts similar to the criteria used in previous analyses of these data~\cite{Ahmed:2008eu,*CDMSScience:2010}. Due to the negligible probability of a WIMP interacting more than once in the apparatus, candidates were required to have energies consistent with noise in all but one detector and have no coincident activity in the plastic scintillator veto. They were further required to have ionization signals consistent with noise in the outer charge electrode. The ionization energy was required to be within $(+1.25,-0.5)$$\sigma$ of the mean ionization energy for nuclear recoils measured from calibration data, which defines the ``nuclear-recoil band.'' This asymmetric band, which has been tightened relative to previous low-energy analyses~\cite{Akerib:2010rr,Ogburn:2008}, was chosen based on calibration data and the observed low-energy backgrounds in the open data in order to maximize sensitivity to nuclear recoils while limiting leakage from electron recoils and zero-charge events.  The recoil-energy range considered for this analysis was 2\textendash100~keV.

\begin{figure}[t]
\centering
\includegraphics[width=3.375in]{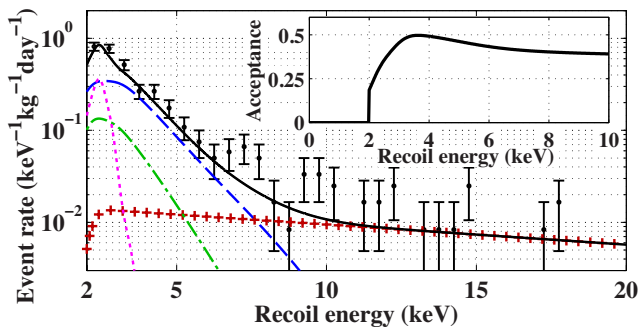}
\caption{(color online). Comparison of the energy spectra for the candidate events and background estimates, co-added over the 8 detectors used in this analysis. The observed event rate (error bars) agrees well with the electron-recoil background estimate (solid), which is a sum of the contributions from zero-charge events (dashed), surface events (+), bulk events (dash-dotted), and the 1.3~keV line (dotted).  The selection efficiencies have been applied to the background estimates for direct comparison with the observed rate, which does not include a correction for the nuclear-recoil acceptance. The inset shows the measured nuclear-recoil acceptance efficiency, averaged over all detectors.}
\label{fig:nr_spec}
\end{figure}

The hardware trigger efficiency was determined using events for which at least one other detector triggered, which provide an unbiased selection of events near threshold. The data are well described by an error function, with the mean trigger threshold varying from 1.5\textendash2.5~keV for the eight Ge detectors. Based on the selection criteria above, the signal acceptance was measured using nuclear recoils from the \cf calibration data.  We calculated the nuclear-recoil band acceptance conservatively by assuming all events with ionization energy $<3\sigma$ above the mean of the distribution were nuclear recoils.  In particular, the zero-charge events described below were included, although their rate in the \cf calibration data is negligible.  The livetime-weighted average of the individual detector selection efficiencies is shown in the inset of Fig.~\ref{fig:nr_spec}, with the largest loss of efficiency coming from the requirement on ionization energy.

\begin{figure}[t]
\centering
\includegraphics[width=3.375in]{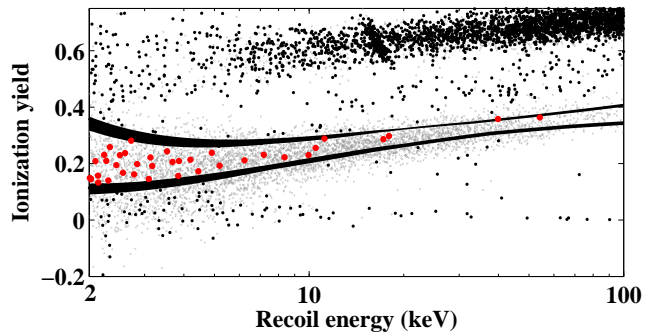}
\caption{(color online). Events in the ionization-yield versus recoil-energy plane for T1Z5. Events within the $(+1.25,-0.5)$$\sigma$ nuclear-recoil band (solid) are WIMP candidates (large dots). Events outside these bands (small, dark dots) pass all selection criteria except the ionization-energy requirement. The widths of the band edges denote variations between data runs.  Events from the \cf calibration data are also shown (small, light dots). The recoil-energy scale assumes the ionization signal is consistent with a nuclear recoil, causing electron recoils to be shifted to higher recoil energies and lower yields.}
\label{fig:cands}
\end{figure}

The energy spectrum for the candidate events passing all selection cuts is shown in Fig.~\ref{fig:nr_spec}. Although the shape of the observed spectrum is consistent with a WIMP signal, we expect that a significant number of the candidates are due to unrejected electron recoils.  Figure~\ref{fig:cands} shows the distribution of candidates in the ionization-yield versus recoil-energy plane for T1Z5. Several populations of events which can leak into the signal region at low energy are apparent.  For each population described below, we measure the rate and energy spectrum in sidebands where the contribution from low-mass WIMPs would be negligible, and extrapolate the observed spectrum to lower energies to estimate the leakage.  The systematic errors introduced by these extrapolations are potentially large and are difficult to quantify. However, as shown in Fig.~\ref{fig:nr_spec} and discussed below, these simple extrapolations can plausibly explain all the observed candidates.

Events with ionization energies consistent with noise are seen below the nuclear-recoil band. Most or all of these ``zero-charge'' events arise from electron recoils near the edge of the detector, where the charge carriers can be completely collected on the cylindrical wall rather than on the readout electrodes. At recoil energies $\gtrsim$10~keV, these events can be rejected using a phonon-based fiducial-volume cut.  At lower energies, reconstruction of the event radius using phonon information is unreliable.  To maintain acceptance of low-energy nuclear recoils, some zero-charge events are not rejected at energies $\lesssim$5~keV where the ionization signal for nuclear recoils becomes comparable to noise. By extrapolating the exponential spectrum observed for zero-charge events above 5~keV, we estimate that they contribute $\sim$50\% of the candidate events.

A second source of misidentified electron recoils comes from events interacting near the detector surfaces, where ionization collection may be incomplete.  These events are primarily concentrated just above the nuclear-recoil band, with an increased fraction leaking into the signal region at low energies.  For recoil energies $\gtrsim$10~keV, nearly all such surface events can be rejected~\cite{Ahmed:2008eu,*CDMSScience:2010} because they have faster-rising phonon pulses than nuclear recoils in the bulk of the detector.  This analysis does not use phonon timing to reject these events since the signal-to-noise is too low for this method to be effective for recoil energies $\lesssim$5~keV. Extrapolating the exponential spectrum of surface events identified above 10~keV implies that $\sim$15\% of the candidates are surface electron recoils. 

At recoil energies $\lesssim$5~keV, the primary ionization-based discrimination breaks down as the ionization signal becomes comparable to noise even for electron recoils with fully collected charge. Extrapolating the roughly constant electron-recoil spectrum observed above 5~keV indicates that $\sim$10\% of the observed candidates arise from leakage of this background into the signal region. Just above threshold, there is an additional contribution to the constant electron-recoil spectrum from the 1.3~keV line, which leaks above the 2~keV analysis threshold since our recoil-energy estimate assumes the ionization signal is consistent with a nuclear recoil. The measured intensity of this line at ionization yields above the signal region indicates that the 1.3~keV line accounts for $\sim$10\% of the observed candidates. T1Z5 has less expected leakage from these fully-collected electron-recoil backgrounds than the average detector since it has the best ionization resolution.

\begin{figure}[t]
\centering
\includegraphics[width=3.375in]{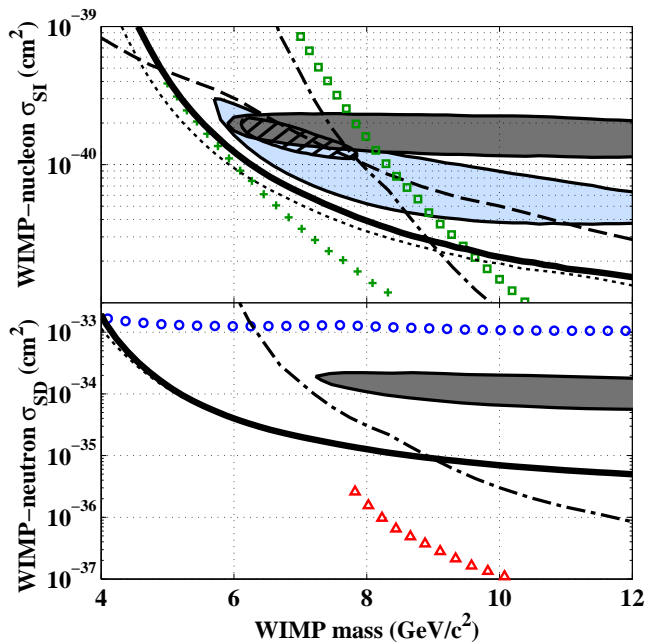}
\caption{(color online). Top: comparison of the spin-independent (SI) exclusion limits from these 
data (solid) to previous results in the same mass range (all at 90\% C.L.). Limits from a low-threshold analysis of the CDMS shallow-site data~\cite{Akerib:2010rr} (dashed), CDMS~II Ge results with a 10~keV threshold~\cite{CDMSScience:2010} (dash-dotted), recalculated for lower WIMP masses, and XENON100 with constant (+) or decreasing ($\square$) scintillation-efficiency extrapolations at low energy~\cite{Aprile:2010xx} are also shown. The filled regions indicate possible signal regions from DAMA/LIBRA~\cite{Bernabei:2008yi,Hooper:2010ly} (dark), CoGeNT (light)~\cite{Aalseth:2010vx,Hooper:2010ly}, and a combined fit to the DAMA/LIBRA and CoGeNT data~\cite{Hooper:2010ly} (hatched). Bottom: comparison of the WIMP-neutron spin-dependent (SD) exclusion limits from these data (solid), CDMS~II Ge results with a 10 keV threshold (dash-dotted), XENON10~\cite{Angle:2008uq} ($\triangle$), and CRESST~\cite{Angloher:2002,*Savage:2004} ($\ocircle$).  The filled region denotes the 99.7\% C.L. DAMA/LIBRA allowed region for neutron-only scattering~\cite{Savage:2008er}.  An escape velocity of 544 km/s was used for the CDMS and XENON100 exclusion limits, whereas the other results assume an escape velocity from 600\textendash650 km/s. Using the same halo parameters as assumed for the allowed regions would lead to slightly stronger limits (dotted).}
\label{fig:lim_plot}
\end{figure}

These estimates indicate that we can claim no evidence for a WIMP signal. However, since the background model involves sufficient extrapolation that systematic errors are difficult to quantify, we do not subtract backgrounds but instead set upper limits on the allowed WIMP-nucleon scattering cross section by conservatively assuming all observed events could be from WIMPs. Limits are calculated using the high statistics version of Yellin's optimum interval method~\cite{Yellin:2002xd,*Yellin:2007bh}. Data from multiple detectors are concatenated as described in~\cite{Akerib:2010rr}. This method allows the choice of the most constraining energy interval on the lowest background detector while applying the appropriate statistical penalty for the freedom to choose this interval. The method and the ordering of detectors by position within the tower were specified with no knowledge of the WIMP candidates to avoid bias. For WIMP masses from 5\textendash8~\gev, the most constraining interval contains events only from T1Z5 and has no dependence on the detector ordering used. The standard halo model described in~\cite{Lewin:1995rx} is used, with specific parameters given in~\cite{Akerib:2010rr,EPAPS}. The candidate event energies and selection efficiencies for each detector are given in~\cite{[{See supplemental material at {\tt http://link.aps.org/\linebreak supplemental/10.1103/PhysRevLett.106.131302}}]EPAPS}.

The limits do not depend strongly on the extrapolation of the ionization yield used at low energies since the Neganov-Luke phonon contribution is small for recoil energies below 4~keV. Conservatively assuming 25\% lower ionization yield near threshold would lead to only $\sim$5\% weaker limits in the 5\textendash10~\gev mass range.

Figure~\ref{fig:lim_plot} (upper panel) shows the resulting 90\% upper confidence limit on the spin-independent WIMP-nucleon scattering cross section.  This analysis provides stronger limits than previous CDMS~II Ge results for WIMP masses below $\sim$9~\gev, and excludes parameter space previously excluded only by the XENON10 and XENON100 experiments for a constant extrapolation of the liquid xenon scintillation response for nuclear recoils below 5~keV~\cite{Aprile:2010xx,Sorensen:2010,Savage:2010dq}.  Our analysis provides stronger constraints than XENON10 and XENON100 below $\sim$7~\gev under conservative assumptions for the scintillation response~\cite{Aprile:2010xx,Hooper:2010ly,Collar:2010gg}.

Spin-dependent limits on the WIMP-neutron cross section are shown in Fig.~\ref{fig:lim_plot} (lower panel), using the form factor from~\cite{Dimitrov:1995}. XENON10 constraints, calculated assuming a constant extrapolation of the scintillation response at low energy~\cite{Angle:2008uq, Collar:2010gg}, are stronger than these results for WIMP masses above $\sim$7~\gev.

These results exclude interpretations of the DAMA/LIBRA annual modulation signal in terms of spin-independent elastic scattering of low-mass WIMPs (e.g.,~\cite{Savage:2010dq,Hooper:2010ly}). We ignore the effect of ion channeling on the DAMA/LIBRA allowed regions since recent analyses indicate channeling should be negligible~\cite{Savage:2010dq,Bozorgnia:2010cr}.  These results are also incompatible with a low-mass WIMP explanation for the low-energy events seen in CoGeNT~\cite{Aalseth:2010vx,Hooper:2010ly}.

The CDMS collaboration gratefully acknowledges the contributions of numerous 
engineers and technicians; we would like to especially thank 
Jim Beaty, Bruce Hines, Larry Novak, 
Richard Schmitt and Astrid Tomada. In addition, we gratefully acknowledge assistance 
from the staff of the Soudan Underground Laboratory and the Minnesota Department of Natural Resources.
This work is supported in part by the 
National Science Foundation (Grant Nos.\ AST-9978911, PHY-0542066, 
PHY-0503729, PHY-0503629, PHY-0503641, PHY-0504224, PHY-0705052, PHY-0801708, PHY-0801712, PHY-0802575, PHY-0847342, and PHY-0855525), by
the Department of Energy (Contracts DE-AC03-76SF00098, DE-FG02-91ER40688, 
DE-FG02-92ER40701, DE-FG03-90ER40569, and DE-FG03-91ER40618), by the Swiss National 
Foundation (SNF Grant No. 20-118119), and by NSERC Canada (Grant SAPIN 341314-07).

\section{Appendix}

The following appendix includes additional details regarding the calibration of the energy scale and expected backgrounds to clarify several of the points raised in~\cite{Collar:2011fk}.

\begin{figure}[th]
\centering
\includegraphics[width=3.375in]{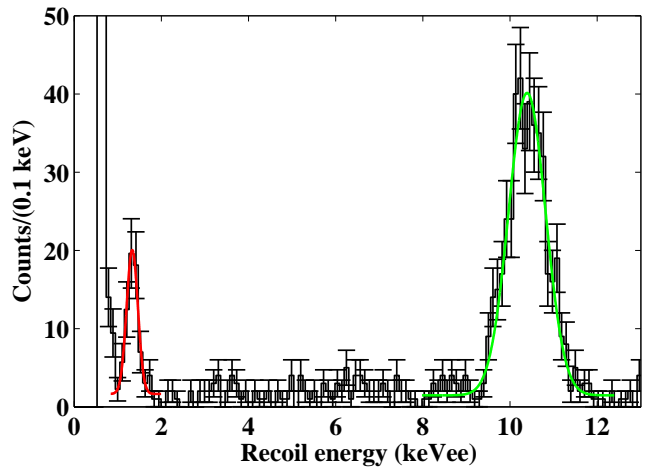}
\caption{Efficiency-corrected phonon recoil-energy spectrum for electron recoils in T1Z5.  The solid lines show fits to the location of the activation lines, which give mean values of 1.333$\pm$0.025 keVee and 10.391$\pm$0.022 keVee. The resolution of the 1.3 keV line is $\sim$100 eVee, consistent with the expected resolution from noise.  The 10.4~keV line is broadened by position dependence of the phonon signal for which no correction has been applied. Cosmogenic $^{65}$Zn has decayed away sufficiently that its contribution to the low energy tail of the activation peaks is negligible~\cite{Ahmed:2009rh}. The relative intensity of the lines measured from the fits is 0.145$\pm$0.030, consistent with expectations.}
\label{fig:act_lines}
\end{figure}

The phonon-based recoil energy scale for this analysis was calibrated using the position of the 1.298 keV and 10.367 keV activation lines which conveniently lie in the recoil energy range of interest for low mass WIMPs.  These lines provide a robust determination of the energy scale for electron recoils near threshold.  We use the most conservative values for the energy scale which are consistent with the position of these activation lines at the 90\% CL for each detector.  Figure~\ref{fig:act_lines} shows the measured positions of the 1.3 keV and 10.4 keV activation lines for T1Z5, after calibrating the energy scale to ensure that the energy is not underestimated for these lines.  The recoil energy units are keVee, or keV ``electron-equivalent'', which indicates that the total phonon signal has been corrected for the contribution from Neganov-Luke phonons assuming that the ionization produced was consistent with an electron recoil.

\begin{figure}[!th]
\centering
\includegraphics[width=3.375in]{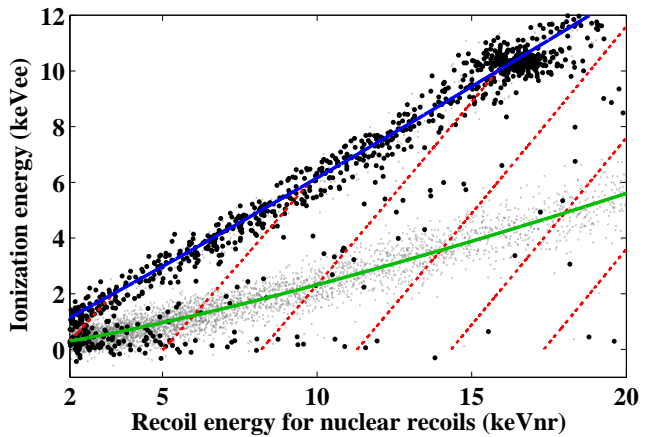}
\caption{Events in the ionization energy vs. recoil energy plane for T1Z5.  Events from the \cf calibration data (small, gray dots) and WIMP search data (large, black dots) are shown.  The recoil energy scale is given by the total phonon energy minus the Neganov-Luke phonon contribution corresponding to the mean ionization for nuclear recoils.  This scale, in units of keVnr, gives the correct recoil energy only for nuclear recoils, while electron recoils appear at higher recoil energy due to the larger contribution of Neganov-Luke phonons.  The solid lines show fits to the mean of the nuclear recoils (green) and electron recoils (blue).  The red dotted curves indicate lines of constant recoil energy when the ionization signal is used to determine the Neganov-Luke phonon contribution. }
\label{fig:nr_band_lin}
\end{figure}

Using the phonon energy scale calibrated with the electron-recoil activation lines, the recoil energy for the candidate nuclear recoils was calculated.  This recoil energy estimate uses only the phonon signal and does not use the ionization signal on an event-by-event basis due to the poorer signal-to-noise of the ionization pulses.  The measured total phonon signal is corrected for the small contribution from Neganov-Luke phonons ($\sim$15\% of the phonon signal for low-energy nuclear recoils) by subtracting the drift heat corresponding to the mean ionization energy measured for nuclear recoils in the \cf calibration data.  Figure~\ref{fig:nr_band_lin} shows the ionization energy versus recoil energy for the WIMP search events and \cf calibration data, using the recoil energy determined by assuming the Neganov-Luke phonon contribution is consistent with a nuclear recoil.

\begin{figure*}[t]
\centering
\includegraphics[width=7in]{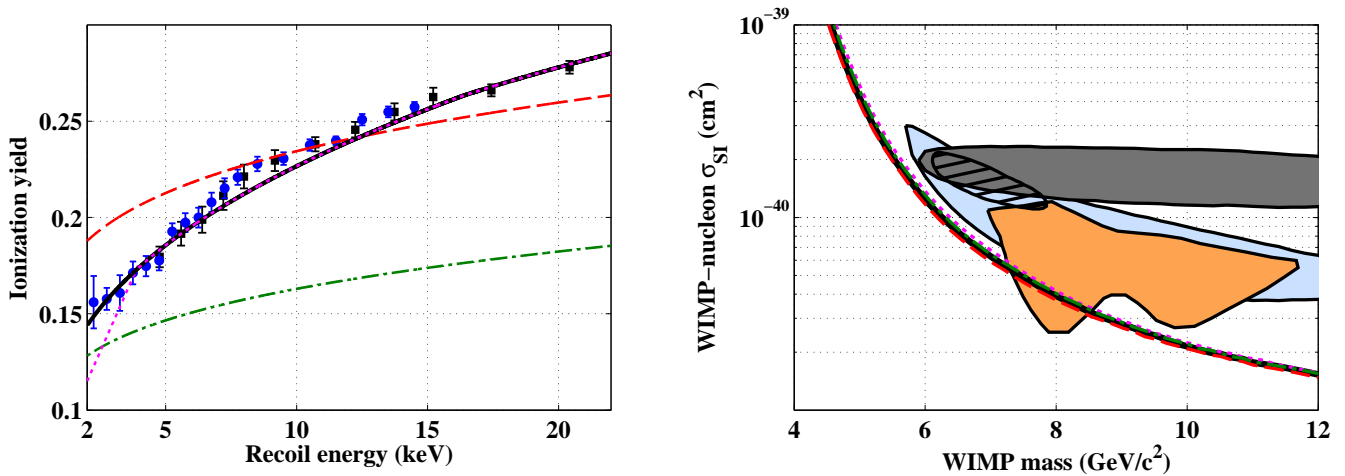}
\caption{ (left) Measurement of the mean ionization yield for nuclear recoils in T1Z5.  Both 3-parameter fits to the yield distribution (black, squares) and 1-parameter fits where the width and amplitude are constrained (blue, circles) are shown.  The black solid line shows the power-law fit to the measured yields above 4~keV, which was extrapolated to lower energies.  This fit was based on data which improperly included events outside the fiducial volume, leading to slightly lower yields than shown by the data points from $\sim$7-15 keV.  This shift has a negligible effect on the energy scale.  Several other models for the ionization yield at low energy are also shown:  Lindhard prediction~\cite{Lindhard:1963,Lewin:1995rx} (red, dashed), steeper yield extrapolation below 4~keV (magenta, dotted), and Lindhard, with $k$=0.1 (green, dot-dashed).  (right) Effect on the exclusion limits from this analysis for the ionization yield models shown on the left.  The limits are only weakly dependent on the extrapolation of the ionization yield at low energies since the Neganov-Luke phonon contribution is small for low-energy nuclear recoils.  The 90\% CL allowed regions for CoGeNT (blue), DAMA/LIBRA (gray), and a combined fit to CoGeNT and DAMA/LIBRA (hatched) from~\cite{Hooper:2010ly} are shown.  The region where models including light WIMPs were found to provide a good fit to CoGeNT in~\cite{Aalseth:2010vx} is also shown (orange).  The limits were calculated using $v_0$=220~km/s and $v_{esc}$=544~km/s, with detailed parameters given in~\cite{EPAPS}.}
\label{fig:band_extrap}
\end{figure*}

To properly correct for the Neganov-Luke phonons, we must determine the ionization yield for nuclear recoils as a function of recoil energy.  Above $\sim$4~keV, depending on detector, the peak of the nuclear recoil ionization distribution lies $>2\sigma$ above noise, and the ionization yield was found by fitting this distribution to a Gaussian in bins of recoil energy.  A power law was then fit to these yield data from $\sim$4-20~keV and extrapolated to determine the ionization yield at lower energies, as shown in Fig.~\ref{fig:band_extrap} (left) for T1Z5.

\begin{figure}[!th]
\centering
\includegraphics[width=3.375in]{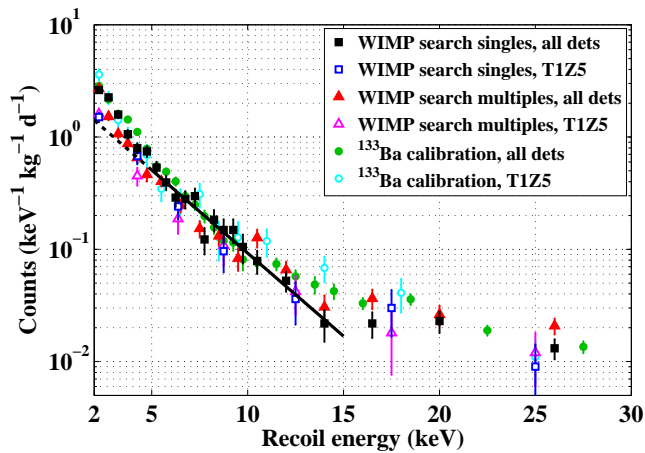}
\caption{Zero-charge event rate vs. recoil energy. The observed rates for single- and multiple-scatter zero-charge events in the WIMP search data are shown, both for the total rate coadded over detectors and for T1Z5 alone.  Below $\sim$5~keV, bulk electron recoils can leak into the zero-charge band, leading to a more rapid increase in the observed spectrum. The rate of zero-charge events in the \ba calibration data is also shown, after subtracting the expected contribution at low energy from the constant bulk electron-recoil spectrum and scaling by a factor of $\sim$$5\times10^{-4}$ to match the zero-charge event rate in the WIMP search data from 5-10~keV.  The solid line shows the exponential fit to the coadded WIMP search singles rate from 5-15~keV, which was extrapolated to estimate the zero-charge event rate near threshold.}
\label{fig:zeroc}
\end{figure}

As a cross-check on this extrapolation, we can extend the fits to lower energies  by fitting only the portion of the distribution which lies $>2\sigma$ above noise.  To ensure that these fits are well-behaved at energies where the peak of the distribution is not contained in the fitting window, we constrain the width of the Gaussian using the measured ionization resolution as a function of energy from activation lines at 1.3, 10.4 and 66.7~keV. The amplitude of the Gaussian is fixed based on Monte Carlo simulations of the \cf neutron calibration recoil-energy spectrum.  We then perform a 1-parameter fit to determine the mean of the distribution.   Both fitting methods give similar yields above $\sim$4~keV, while the fixed width and amplitude fits can be extended down to the threshold of 2~keV.  The yields determined by the 1-parameter fits agree well with the power-law extrapolation.  Details of the measurement of the ionization yield will be presented in an upcoming publication~\cite{Ahmed:2011}.  

Figure~\ref{fig:band_extrap} (right), shows that the energy scale and limits are only weakly dependent on the extrapolation of the ionization yield at low energy since the Neganov-Luke phonons contribute only a small fraction of the total phonon signal.  Conservatively assuming ionization yields which are $\sim$25\% lower than those suggested by our data, and by previous measurements by other groups in this energy range (see e.g.~\cite{Hooper:2010ly} and references therein), would not change the conclusions of this analysis.

\begin{figure*}[!th]
\centering
\includegraphics[width=7in]{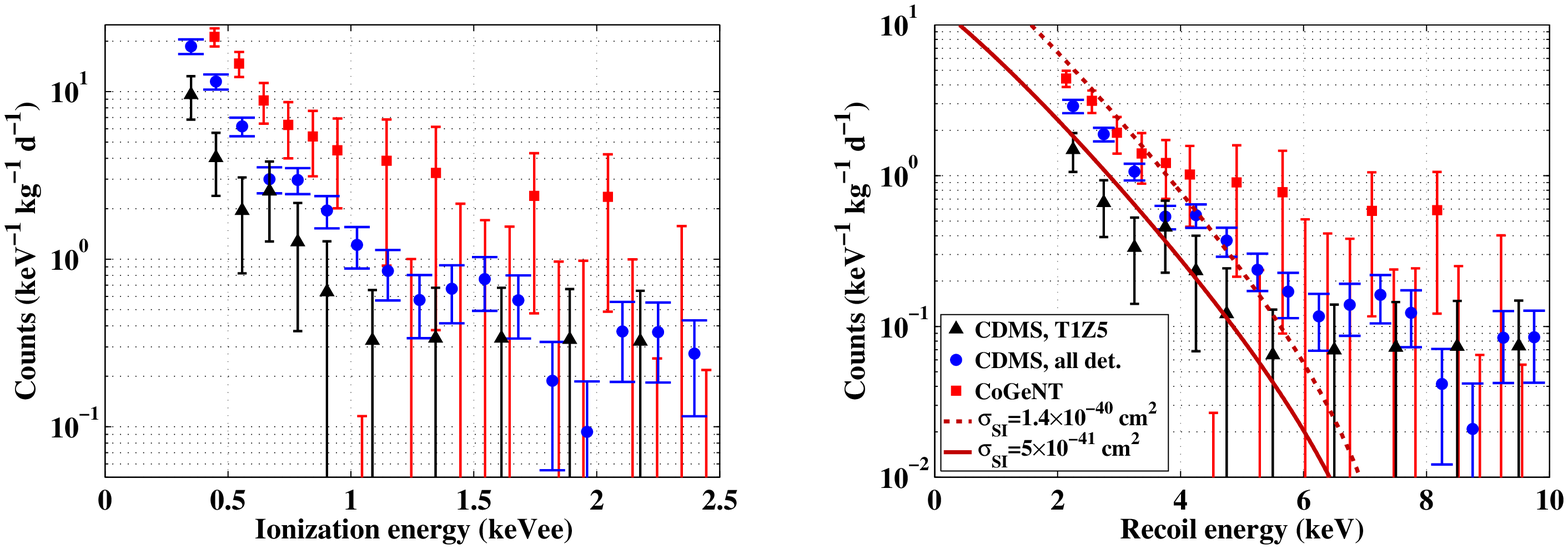}
\caption{ (left) Comparison of the observed rate in CoGeNT (red, squares), CDMS coadded over all detectors (blue, circles), and CDMS detector T1Z5  (black, triangles), which sets the strongest constraints in the 5-10~\gev mass range.  All data are corrected for the nuclear-recoil acceptance efficiencies.  The CoGeNT data shown have the L-shell electron capture peaks and a constant background subtracted~\cite{Collar:2011fk}.  The CDMS ionization spectra are based on the recoil energy reconstructed from the phonon signal alone and have been converted to ionization energy using the ionization yields shown as the solid black curve in Fig~\ref{fig:band_extrap}.   (right) Comparison of the same data versus recoil energy.  The CoGeNT data have been converted to recoil energy using the quenching factor assumed in~\cite{Hooper:2010ly}.  This quenching factor is slightly higher than the ionization yields measured by CDMS, causing the spectra to appear more compatible than when plotted versus ionization energy.  The recoil spectra for the WIMP models considered in Fig.~1 of ref.~\cite{Collar:2011fk} are also shown. The dotted line indicates the expected spectrum for $m_\chi$=7~\gev and $\sigma_{SI}$=$1.4\times10^{-40}$~cm$^2$, corresponding to a WIMP model from a simultaneous fit to the DAMA/LIBRA and CoGeNT data~\cite{Hooper:2010ly,Collar:2011fk}.  The solid line shows the spectrum for the same WIMP mass and $\sigma_{SI}$=$5\times10^{-41}$~cm$^2$, which is described as the best fit for a $m_\chi$=7~\gev WIMP to the CoGeNT data in~\cite{Collar:2011fk}.  This point lies outside the CoGeNT allowed regions from~\cite{Hooper:2010ly,Aalseth:2010vx} and is not excluded by our analysis.  The WIMP spectra assume $v_0$=220~km/s and $v_{esc}$=544~km/s.}
\label{fig:spec_comp}
\end{figure*}

Although the correction for the Neganov-Luke phonons does not contribute significant uncertainty to the recoil energy scale, larger errors are possible if the phonon collection for nuclear recoils differs from that for electron recoils.  The absolute nuclear recoil energy scale is constrained by the measured ionization yields shown in Fig.~\ref{fig:band_extrap}, which are $\sim$15\% lower than previous measurements in the energy range of interest.  Assuming that the ionization produced is consistent with previous measurements of the quenching factor, the measured yields imply an overestimate of the recoil energy scale near threshold.  To be conservative, we do not apply a correction to the recoil energy scale based on the measured yields, but such a correction would improve our constraints in the 5-10~\gev mass range.

We calculated limits on the WIMP-nucleon cross section by conservatively assuming all the candidate events could arise from WIMP-induced nuclear recoils.  However, we expect significant backgrounds at low energy which can mimic a WIMP signal.  Although estimates of these backgrounds do not affect the limits on the WIMP-nucleon cross section derived, they indicate that we cannot claim evidence for a WIMP signal since all observed candidates can plausibly be accounted for by expected backgrounds.

In particular, zero-charge events are expected to be the limiting background in the energy range of interest for WIMPs with masses $\sim$7~\gev.  These events are consistent with electron recoils at very high radius in the detector, where the ionization can be completely collected on the cylindrical walls instead of on the ionization electrodes.  Zero-charge events are seen in the WIMP search data for both events interacting in a single detector (``singles'') and events with energy deposited in more than one detector (``multiples'').  As shown in Fig.~\ref{fig:zeroc}, the zero-charge multiples and singles are observed to occur with similar rates and approximately exponential spectra below $\sim$15~keV.  Due to the negligible probability of a WIMP interacting in more than one detector, the multiple-scatter zero-charge events can arise only from backgrounds.  We expect a corresponding contribution to the zero-charge singles rate from the same background sources when energy is deposited in inactive portions of the experiment and only a single detector.  For bulk electron recoils, the rate of tagging multiple scatters is approximately twice the single-scatter rate.  For events interacting on the cylindrical surfaces of the detectors, the multiple-scatter tagging is expected to be less efficient since the cylindrical walls face the copper detector housings rather than another detector. 

In addition, zero-charge events are seen in the \ba calibration data with a similar spectrum to the WIMP search data.  Their rate in the \ba calibration data is significantly higher, with $\sim$4 times more zero-charge events observed from 5-10~keV in an exposure which was $\sim$1/500 as large.  This demonstrates that electron recoils can produce zero-charge events with an increasing spectrum at low energy.  Although the low-energy gamma spectrum is approximately constant, we expect that zero-charge events should increase in rate at low energy since only those events which interact in a small dead layer near the cylindrical wall will have their charge trapped and pass the fiducial volume selection.  Due to the rapidly decreasing penetration depth of low-energy external gammas, an increasing fraction of low-energy events will interact in this dead layer.  However, the zero-charge event rate does not scale directly with the electron-recoil rate since only $\sim$1/6 as many zero-charge events were observed in the \ba calibration data as in the WIMP search singles and multiples from 5-10~keV, when normalized to the total number of bulk electron recoils.

To estimate the contribution of zero-charge events at low energy, we extrapolated the approximately exponential spectrum observed for the coadded WIMP search singles from 5-15~keV to lower energies, as shown in Fig.~\ref{fig:zeroc}.  This extrapolation was found to be relatively insensitive to the choice of lower boundary for the fit, with a 20\% variation in the expected rate from 2-5~keV when the lower edge of the fitting window was varied from 5-8~keV.  Near 5~keV, there is some overlap of the nuclear-recoil and zero-charge distributions.  If WIMP-induced nuclear recoils are present, a small fraction could appear in the zero-charge band and bias the background estimate.  Even conservatively assuming that all events in the (+1.25,-0.5)$\sigma$ nuclear-recoil acceptance region are due to WIMP-induced nuclear recoils, we would expect only $\sim$15\% of the observed single-scatter zero-charge events from 5-6~keV would be due to WIMPs.  The overlap of the distributions decreases rapidly with energy, and any error introduced is small relative to possible errors from the extrapolation itself.   Although these estimates indicate that we expect a significant zero-charge contribution in the region of interest for light WIMPs, we do not subtract this background due to possibly significant systematic errors introduced by the extrapolation which are difficult to quantify.

A direct comparison of the CDMS and CoGeNT spectra is shown in Fig.~\ref{fig:spec_comp}.  Both the CDMS data coadded over all detectors and the data for T1Z5 alone are shown.  T1Z5 has the best ionization resolution, leading to the best discrimination against electron-recoil backgrounds, and it dominates the sensitivity of this analysis in the 5-10~\gev mass range.  Even with no background subtraction, the CDMS data for T1Z5 are incompatible with a low-mass WIMP interpretation for the entire CoGeNT excess.  Figure~\ref{fig:spec_comp} also shows the expected WIMP spectra for the models considered in~\cite{Collar:2011fk}.  The solid line shows the WIMP model described as the best fit to the CoGeNT data for $m_\chi$=7~\gev.  As shown in Fig.~\ref{fig:spec_comp}, this model requires a significant fraction of the CoGeNT low-energy excess to be due to an exponential background, and it is not excluded by this analysis.  To make the observed spectra for CDMS and CoGeNT compatible would require the majority of the CoGeNT excess to be due to low-energy backgrounds and the CDMS backgrounds to be smaller than expected.

\bibliography{c38_LowT}
\bibliographystyle{apsrev4-1}

\end{document}